# P2P Appliance Calculation Method For Trust between Nodes within a P2P Network

Adis Medic[1], Adis Golubovic[2]

[1] InfoSys LTD,
Kolodvorska bb, 77240 Bosanska Krupa, Una Sana Canton, Bosnia and Herzegovina
Ceravačka brda 106, 77000 Bihać, Una Sana Canton, Bosna and Herzegovina

[2] Pimary School „Podzvizd", Velika Kladusa, Bosnia and Herzegovina,

**Abstract**
Modern ways of communications, such as in a web services environment, also influences trust relationships between organisations. This concept of web-based (way towards semantic web services) trust is new and has as yet not been resolved. We hope that some of the trust properties mentioned above can be successfully employed to improve the understanding of trust between machines. So, trust is a vital ingredient of any successful interaction between individuals, among organizations and/or in society at large. In this paper, we suggested a trust model using fuzzy logic in semantic network of nodes. Trust is an aggregation of consensus given a set of past interaction among nodes (semantic network based on machines, agents etc.). We applied our suggested model to semantic networks in order to show how trust mechanisms are involved in communicating algorithm to choose the proper path from source to destination. Authors use the terms untrust and distrust as synonyms for the condition opposite to the trust.

## 1. Introduction

Trust relationships between organisations are, among others, influenced by culture and adherence to codes of best practices. A model of inter-organisational trust illustrates that trust is dependent on: competence, consistent positive behaviours and goodwill [14]. Trust is a central component of the Semantic Web vision. The Semantic Web stack [3][4][10] has included all along a trust layer to assimilate the ontology, rules, logic, and proof layers. Trust often refers to mechanisms to verify that the source of information is really who the source claims to be. Signatures and encryption mechanisms should allow any consumer of information to check the sources of that information. In addition, proofs should provide a tractable way to verify that a claim is valid. In this sense, any information provider should be able to supply upon request a proof that can be easily checked that certifies the origins of the information, rather than expect consumers to have to generate those proofs themselves through a computationally expensive process. The web motto "Anyone can say anything about anything" makes the web a unique source of information, but we need to be able to understand where we are placing our trust [1)][2][3][4].

Trust plays a central role in many aspects of computing, especially those related to network use. Whether downloading and installing software, buying a product from a web site, or just surfing the Web, an individual is faced with trust issues. Does this piece of software really do what it says it does? Trust has another important role in the Semantic Web, as agents and automated reasoners need to make trust judgements when alternative sources of information are available [8]. Computers will have the challenge to make judgements in light of the varying quality and truth that these diverse "open" (unedited, uncensored) sources offer. Today, web users make judgments routinely about which sources to rely on since there are often numerous sources relevant to a given query, ranging from institutional to personal, from government to private citizen, from objective report to editorial opinion, etc. These trust judgements are made by humans based on their prior knowledge about a source's perceived reputation, or past personal experience about its quality relative to other alternative sources they may consider. Humans also bring to bear vast amounts of knowledge about the world they live in and the humans that populate the web with information about it. In more formal settings, such as e-commerce and e-science, similar judgments are also made with respect to publicly available data and services. All of these important trust judgments are currently in the hands of humans. This will not be possible in the Semantic Web, where humans will not be the only consumers of information. Agents will need to automatically make trust judgments to choose a service or information source while performing a task [6].

Reasoners will need to judge which of the many information sources available, at times contradicting one another, are more adequate for answering a question. In a Semantic Web where content will be reflected in ontologies and axioms, how will a computer decide what sources to trust when they offer contradictory information? What mechanisms will enable agents and reasoners to make trust judgments in the Semantic Web? Trust is not a new research topic in computer science, spanning areas as diverse as security and access control in computer networks, reliability in distributed systems, game theory and agent systems, and policies for decision making under uncertainty. The concept of trust in these different communities varies in how it is represented, computed, and used. While trust in the Semantic Web presents unique challenges [13], prior work in these areas is relevant and should be the basis for future research.

Trust can be viewed at a micro or macro level. At the micro level, a series of tactics can, in various circumstances, help create or preserve trust. At the macro level, such tactics need to be combined into trust strategies. Various tactics were set out, some of which are variants on others. For example, there are many variations on the tactic of restricting those sources of knowledge that a knowledge technology uses, including relying on branded websites, and demanding verifiable certification of provenance. Managing trust is a key managerial requirement for the semantic web, and an interesting demand that has come to light is for informative metadata about knowledge sources that can be used for assessing trustworthiness [11].

## 2. Modeling a Fuzzy and Mathematical model

While a lot of concepts as well as practical applications for the lower layers (Fig. 1) of the Semantic Web exist and layers like "Logic" and "Proof" are not necessary in a lot of e-commerce use-cases, the top-layered "Trust" concept has still been far away from implementations during the last years.

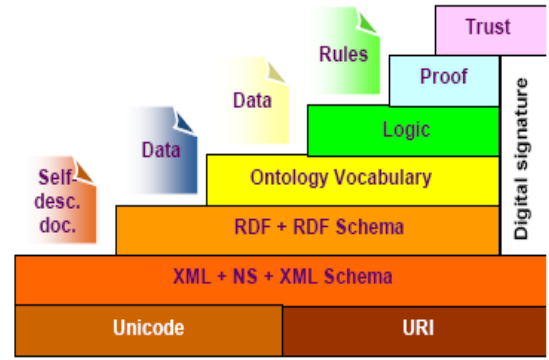

Fig. 1. Stack for the semantic web [7][10]

Within this approach a link to define trustful Semantic-Web-data of a company is integrated. Similar projects for private usage map this approach to the area of social networking platforms like Facebook[1]. The basic idea is to provide an easy method for web users to indicate data within the web as trustful, so that intelligent web applications can work with this information without any further trust proof mechanisms like digital signatures[15]. Trust is one of the major problems for the success of computer supported society, smart physical environment, virtual reality, virtual organization, computer mediated interaction etc. It seems important to study people's trust in the computational infrastructure, people's trust in potential partners, information sources, data, mediating agents, personal assistants and agents' trust in other agents and processes. Trust is indeed a problem: for example, in e-commerce it is far from obvious whether existing paper-based techniques for fraud detection and prevention are adequate to establish trust in an electronic network environment, where you usually never meet your trade partner face to face, and where messages can be read or copied a million times without leaving any trace.

Of course, the notion of trust is also important in other domains of agents' theory, beyond that of electronic commerce. For example, trust is relevant in human-computer interaction, e.g. the trust relation between the user and his personal assistant (and, in general, his computer). It is also critical for modeling and supporting groups and teams, organizations, co-ordination, negotiation through computational devices, with the related trade-off between local/individual utility and global/collective interest; or even in modeling distributed knowledge and its circulation.

---
[1] See http://opentrust-project.com for details.

In conclusion, the notion of trust is crucial for all the major topics of Multi-Agent systems. In all these contexts, different kind of trust are needed and should be modeled and supported:

- trust in the environment and infrastructure (the socio-technical system);
- trust in your agent and in mediating agents;
- trust in the potential partners;
- Trust in the warrantors and authorities (if any)[1].

The problem is therefore how to build trust in users and agents and how to maintain it. Security measures are not enough, interactivity and knowledge are not enough. Building trust in fact is not just a matter of protocols, architectures, mind-design, clear rules and constraints, controls and guarantees. Trust in part is a socially emergent phenomenon; it is a mental stuff yet it is grounded in socially situated agents and it is based on social context.

In this paper authors will show how to calculate a path(s) and trustfulness of nodes inside the paths. These paths are channels for data interchange among and inside various networks and intelligent systems or Semantic web repositories [16]. Also, it will be presented mathematical and fuzzy model and formulae for trust factor calculation, explained on examples. First of all, authors will explain an situation that worked for analysis and show primitive and simplified path-route modeling started scratch, also it will be shown a step-by-step calculation for path trust factor and confirmation of these calculation by proposed mathematical model.

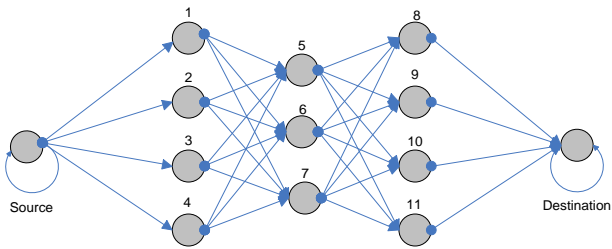

Fig. 2. P2P network model and possible packet routes

Fig. 2 shows the initial model of tested network on which to perform research of confidentiality between the nodes. Values for Trust/Untrust[9] are assumed values. In this paper authors will deal with the method of measurement, but authors propose a new method for the most confidential way (for packet or query results or some other traffic) through the network if we are familiar with the value. On next few figures authors will present values for individual nodes, trust and distrust between the nodes in the network:

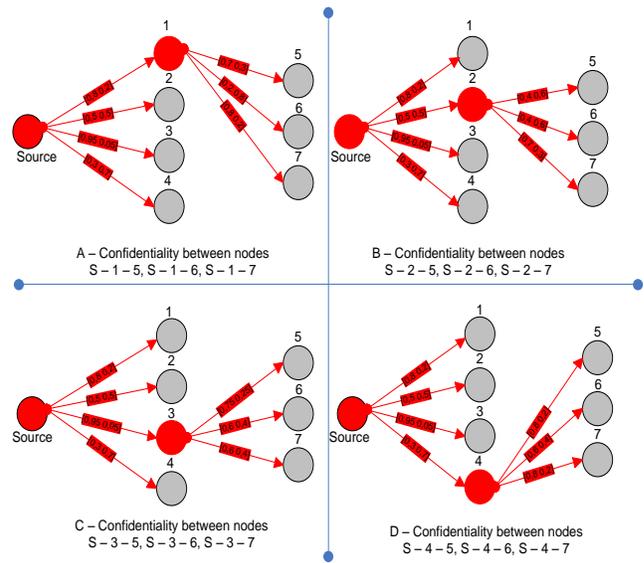

Fig. 3. Assumed Trust/Untrust values between nodes (from Source)

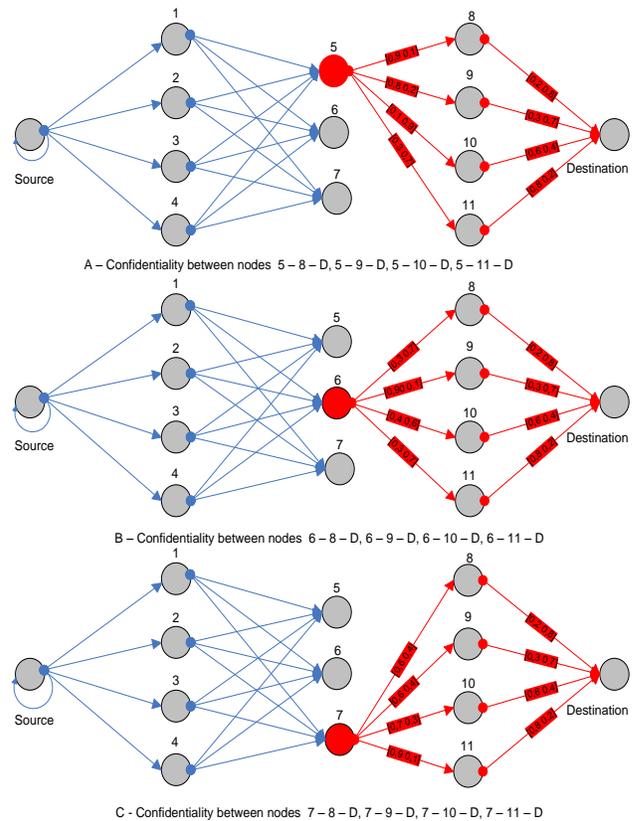

Fig. 4. Assumed Trust/Untrust values between nodes (to Destination)

---
[1]See http://www.istc.cnr.it/T3/ for details

Table 1. Trust Factor Scale

| Trusted | Untrusted | |
|---|---|---|
| 1 | 0 | Very high (VH) |
| 0,85 | 0,15 | |
| 0,70 | 0,30 | High (H) |
| 0,50 | 0,50 | Indifferent (I) |
| 0,30 | 0,70 | Low (L) |
| 0 | 1 | Very low (VL) |

Table 2. Possible Packet Route (from Figure 2)

| | | | |
|---|---|---|---|
| $P_1 =$ S→1→5→8→D | $P_{13} =$ S→2→5→8→D | $P_{25} =$ S→3→5→8→D | $P_{37} =$ S→4→5→8→D |
| $P_2 =$ S→1→5→9→D | $P_{14} =$ S→2→5→9→D | $P_{26} =$ S→3→5→9→D | $P_{38} =$ S→4→5→9→D |
| $P_3 =$ S→1→5→10→D | $P_{15} =$ S→2→5→10→D | $P_{27} =$ S→3→5→10→D | $P_{39} =$ S→4→5→10→D |
| $P_4 =$ S→1→5→11→D | $P_{16} =$ S→2→5→11→D | $P_{28} =$ S→3→5→11→D | $P_{40} =$ S→4→5→11→D |
| $P_5 =$ S→1→6→8→D | $P_{17} =$ S→2→6→8→D | $P_{29} =$ S→3→6→8→D | $P_{41} =$ S→4→6→8→D |
| $P_6 =$ S→1→6→9→D | $P_{18} =$ S→2→6→9→D | $P_{30} =$ S→3→6→9→D | $P_{42} =$ S→4→6→9→D |
| $P_7 =$ S→1→6→10→D | $P_{19} =$ S→2→6→10→D | $P_{31} =$ S→3→6→10→D | $P_{43} =$ S→4→6→10→D |
| $P_8 =$ S→1→6→11→D | $P_{20} =$ S→2→6→11→D | $P_{32} =$ S→3→6→11→D | $P_{44} =$ S→4→6→11→D |
| $P_9 =$ S→1→7→8→D | $P_{21} =$ S→2→7→8→D | $P_{33} =$ S→3→7→8→D | $P_{45} =$ S→4→7→8→D |
| $P_{10} =$ S→1→7→9→D | $P_{22} =$ S→2→7→9→D | $P_{34} =$ S→3→7→9→D | $P_{46} =$ S→4→7→9→D |
| $P_{11} =$ S→1→7→10→D | $P_{23} =$ S→2→7→10→D | $P_{35} =$ S→3→7→10→D | $P_{47} =$ S→4→7→10→D |
| $P_{12} =$ S→1→7→11→D | $P_{24} =$ S→2→7→11→D | $P_{36} =$ S→3→7→11→D | $P_{48} =$ S→4→7→11→D |

Trust factor is presented by formula:

$$F = [T\ U], \quad (1)$$

Where:
T means Trust factor, and
U is Untrust factor (nonT), so we have the following:

- The values of confidentiality are possible in the following intervals:

$$F = [T, \neg T] \text{ or } F=[T, U], \text{ where } \neg T=U \quad (2)$$

the extreme factor values

$$F_{max} = [1,0] \text{ or } F_{max} = [100\%, 0\%], \quad (3)$$

For the highest Trust value (or the lowest Untrust value)

$$F_{min} = [0,1] \text{ or } F_{min} = [0\%, 100\%], \quad (4)$$

For the highest Untrust value (or the lowest Trust value).

Maximum confidentiality has an initial node S,

$$F = [1,0] \quad (5)$$

$$F = [T_n\ U_n] \cdot \begin{bmatrix} T_{min} & U_{n+1} \\ T_{max} & T_{indif} \end{bmatrix} = [F_T\ F_U] \quad (6)$$

The Trust factor is analyzed by following logic-algorithm assumptions:

If

$$F_T > F_U \text{ then, Node } n+1 \text{ is acceptable} \quad (7)$$

If

$$F_T < F_U \text{ then Node } n+1 \text{ is not acceptable} \quad (8)$$

Else,

Node $n+1$ indifferent acceptance $\quad (9)$

## 3. Calculation of paths and path table

Confidentiality testing for nodes in a $P_{36}$ path (Test of confidentiality nodes in a path that represents the most likely route for packets from source to the destination – Figure 5):

S→3→7→11→D

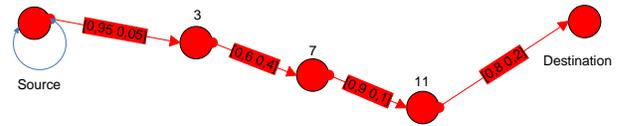

Fig. 5. The most likely packet route, from S(ource) to D(estination)

$$S \to 3 = [1\ 0] \cdot \begin{bmatrix} 0,51 & 0,05 \\ 1,00 & 0,50 \end{bmatrix} = [(1 \cdot 0,51 + 0 \cdot 1)(1 \cdot 0,05 + 0 \cdot 0,5)] = [0,51\ 0,05] \quad (10)$$

$$3 \to 7 = [0,95\ 0,05] \cdot \begin{bmatrix} 0,51 & 0,40 \\ 1,00 & 0,50 \end{bmatrix} = [(0,95 \cdot 0,51 + 0,05 \cdot 1)(0,95 \cdot 0,4 + 0,05 \cdot 0,5)] = [0,53\ 0,40] \quad (11)$$

$$7 \to 11 = [0,6\ 0,4] \cdot \begin{bmatrix} 0,51 & 0,10 \\ 1,00 & 0,50 \end{bmatrix} = [(0,6 \cdot 0,51 + 0,4 \cdot 1)(0,6 \cdot 0,1 + 0,4 \cdot 0,5)] = [0,706\ 0,26] \quad (12)$$

$$11 \to D = [0,9\ 0,1] \cdot \begin{bmatrix} 0,51 & 0,20 \\ 1,00 & 0,50 \end{bmatrix} = [(0,9 \cdot 0,51 + 0 \cdot 1)(0,9 \cdot 0,2 + 0,1 \cdot 0,5)] = [0,559\ 0,23] \quad (13)$$

Testing untrust of nodes in a path $P_{36}$ (Test for untrust of nodes in a path that represents the most likely

route for packets from source to the destination – Figure 5):

S→3→7→11→D:

$S \rightarrow 3 = [0\ 1] \cdot \begin{bmatrix} 0,49 & 0,95 \\ 0,00 & 0,50 \end{bmatrix} = [(0 \cdot 0,49 + 1 \cdot 0)(0 \cdot 0,95 + 1 \cdot 0,5)] = [0\ 0,50]$ (14)

$3 \rightarrow 7 = [0,05\ 0,95] \cdot \begin{bmatrix} 0,49 & 0,60 \\ 0,00 & 0,50 \end{bmatrix} = [(0,05 \cdot 0,49 + 0,95 \cdot 0)(0,05 \cdot 0,6 + 0,95 \cdot 0,5)] = [0,024\ 0,505]$ (15)

$7 \rightarrow 11 = [0,4\ 0,6] \cdot \begin{bmatrix} 0,49 & 0,90 \\ 0,00 & 0,50 \end{bmatrix} = [(0,4 \cdot 0,49 + 0,6 \cdot 0)(0,4 \cdot 0,9 + 0,6 \cdot 0,5)] = [0,196\ 0,66]$ (16)

$11 \rightarrow D = [0,1\ 0,9] \cdot \begin{bmatrix} 0,49 & 0,80 \\ 0,00 & 0,50 \end{bmatrix} = [(0,1 \cdot 0,49 + 0,9 \cdot 0)(0,1 \cdot 0,8 + 0,9 \cdot 0,5)] = [0,049\ 0,53]$ (17)

The highest Trust factor is on following paths:
$P_{12}$ = S→1→7→11→D (18)
$\left(\dfrac{0,8 + 0,8 + 0,9 + 0,8}{4}\right) = 0,825 => 0,82$
$P_{36}$ = S→3→7→11→D (19)
$\left(\dfrac{0,95 + 0,6 + 0,9 + 0,8}{4}\right) = 0,8125 => 0,81$

The lowest Untrust factor, shows that this path has the highest Trust factor. And here are the most reliable paths:
$P_{12}$ = S→1→7→11→D (20)
$\left(\dfrac{0,2 + 0,2 + 0,1 + 0,2}{4}\right) = 0,175 => 0,17$
$P_{36}$ = S→3→7→11→D (21)
$\left(\dfrac{0,05 + 0,4 + 0,1 + 0,2}{4}\right) = 0,1875 => 0,18$

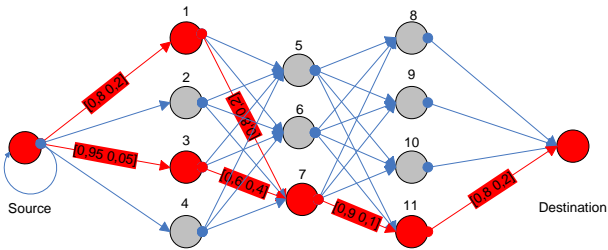

Fig. 6. The most likely packet route through the P2P network

Important (very high) Trust value for paths:
$P_{36}$ = S→3→7→11→D (22)
$\left(\dfrac{0,95 + 0,6 + 0,9 + 0,8}{4}\right) = 0,8125 => 0,81$
$P_{12}$ = S→1→7→11→D (23)
$\left(\dfrac{0,8 + 0,8 + 0,9 + 0,8}{4}\right) = 0,825 => 0,82$

Also, important (very small) Trust value (or very high Untrust value) for paths (it automatically presents a calculation check for previous results for Trust value):
$P_{36}$ = S→3→7→11→D (24)
$\left(\dfrac{0,05 + 0,4 + 0,1 + 0,2}{4}\right) = 0,1875 => 0,18$
$P_{12}$ = S→1→7→11→D (25)
$\left(\dfrac{0,2 + 0,2 + 0,1 + 0,2}{4}\right) = 0,175 => 0,17$

By analyzing previous calculation, we can conclude that the highest total confidentiality is the path P12, but most likely route packets will take place, is P36 path because node S has the highest confidentiality towards third node.

## 4. Conclusion and further work

Note that during the analysis was used mesh topology, and we consistently follow the P2P communication appliance [12] among nodes within the network. Also, the authors approved a proposed way of calculating the importance of certain nodes within the path, which is considered to be acceptable for realization of communication and exchange of knowledge or information and in order to achieve results.

In future research, it will be very important, also challenging, to establish accurate and relevant metrics model. It is necessary to try to optimize either combination of different metrics to get meta-model for suitable metrics [12][5] for use or adapted for use in the previously shown method. The authors will, in future work, try to propose a fuzzy routing table with some interesting factors which directly involve with confidentiality of trusted nodes, such are factor for node that has unreliable neighbor nodes, factor for node that is trusted but not used for packet transfer, or factor for node that transferred a packet. By using fuzzy logic to determine the weights for direct trust as well as reputation, our fuzzy trust model becomes flexible to rely on direct trust or on reputation based trust.

**Adis Medić** received his masters degree (Dipl.-Ing. EE-Inf/ MScEE-Inf; 2008) in Electrotechnics – Informatics from the Technical Faculty of Bihać, University of Bihać. He is a PhD student on Faculty of Organization and Informatics, University of Zagreb. He is currently employed at Infosys ltd as a system and network engineer.

**Adis Golubović** received his masters degree (Dipl.-Ing. EE-Inf/ MScEE-Inf; 2009) in Electrotechnics – Informatics from the Technical Faculty of Bihać, University of Bihać. He is a PhD student on Faculty of Organization and Informatics, University of Zagreb. He is teaching Informatics at local primary school.